\documentclass[]{aastex631}
\usepackage{float}
\usepackage[T1]{fontenc} % if needed
\usepackage{amssymb}
\usepackage{bm}
\usepackage{amsfonts}
\usepackage{latexsym}
\usepackage[latin1]{inputenc}
\usepackage{graphicx}
\usepackage{amsmath}
\usepackage{mathpazo}
\usepackage{amsmath}
\usepackage{mathrsfs}
\usepackage{amsthm}
\usepackage{graphicx}
\usepackage{hyperref}
\usepackage{enumitem}
\usepackage{textcomp}
\usepackage{booktabs}
\usepackage{subfigure}
\usepackage{dcolumn}
\usepackage{ragged2e}
\usepackage{hyperref}
\usepackage{footmisc}

\allowdisplaybreaks[1]
\newcommand{\m}{\mathring}
\renewcommand{\arraystretch}{1.1}
\begin{document}

\title{Stability of \texorpdfstring{$f(Q, B)$}{} Gravity via Dynamical System Approach: a Comprehensive Bayesian Statistical Analysis}

\author[0000-0001-5934-3428]{Santosh V. Lohakare}
\affiliation{Department of Mathematics,
Birla Institute of Technology and Science-Pilani, Hyderabad Campus, Hyderabad-500078, India.}

\author[0000-0001-5527-3565]{B. Mishra}
\affiliation{Department of Mathematics,
Birla Institute of Technology and Science-Pilani, Hyderabad Campus, Hyderabad-500078, India.}



\begin{abstract}

In this work, we explore the cosmological stability of $f(Q, B)$ gravity using a dynamical system approach, where $Q$ denotes the nonmetricity scalar and $B$ represents the boundary term. We determine the model parameters of $f(Q, B)$ through Bayesian statistical analysis, employing Markov Chain Monte Carlo techniques. This analysis incorporates numerical solutions and observational data from cosmic chronometers, the extended Pantheon$^+$ data set, and baryonic acoustic oscillation measurements. Our findings reveal a stable critical point within the dynamical system of the model, corresponding to the de Sitter phase, which is consistent with current observations of the Universe dominated by dark energy and undergoing late-time accelerated expansion. Additionally, we utilize Center Manifold Theory to examine the stability of this critical point, providing deeper insights into the behavior of the model. The cosmological implications of $f(Q, B)$ gravity indicate a smooth transition in the deceleration parameters from deceleration to the acceleration phase, underscoring the potential of the model to describe the evolution of the Universe. Our results suggests that the $f(Q, B)$ model presents a viable alternative to the standard $\Lambda$CDM model, effectively capturing the observed acceleration of the Universe and offering a robust framework for explaining the dynamics of cosmic expansion.
\end{abstract}



\section{Introduction} \label{Sec 1}

    The Lambda Cold Dark Matter ($\Lambda$CDM) model is widely regarded as the standard cosmological model, providing a fundamental framework for interpreting cosmic microwave background (CMB) data \citep{Aghanim_2018_641}. It incorporates cold dark matter, and late-time accelerated expansion driven by dark energy (DE) within a homogeneous and isotropic Universe. The cosmological constant $(\Lambda)$, integrated into Einstein's field equations, has a robust theoretical basis. However, recent observations have highlighted discrepancies and tensions in $\Lambda$CDM model parameters across different cosmic epochs \citep{DiValentino_2021_38, Perivolaropoulos_2022_95, Lohakare_2024_MNRAS_FG}, alongside persistent theoretical challenges associated with $\Lambda$ \citep{Weinberg_1989_61, Peebles_2003_75}. These inconsistencies suggest that the $\Lambda$CDM model may not fully account for all missing physics. The discovery of the accelerating Universe expansion is one of the most surprising findings in cosmology, indicating either a breakdown of Einstein's theory of gravity on cosmological scales or the dominance of DE with exotic properties. Cosmic acceleration has driven efforts to uncover its origin, with experiments measuring expansion and growth to one percent precision or better. General relativity (GR) is non-renormalizable at microscopic scales \citep{Hooft_1974_20, Goroff_1985_160}, and the absence of a coherent quantum gravity theory necessitates modifications or extensions of GR \citep{Hooft_1974_20, Gleiser_2005_58}. The standard approach develops gravity theories that encompass GR in specific scenarios while introducing additional degrees of freedom \citep{Nojiri_2011_505, Capozziello_2011_509, Bahamonde_2023_86, Franco_2020_80, Heisenberg_2023_83}. These theories arise from fundamental considerations of gravitational field structures and principles \citep{Bohmer_2021_104, Bohmer_2023_64, CANTATA_2023_2105.12582}, with a focus on maintaining the principle of equivalence at quantum scales and addressing causal and geodesic structures. The metric requirement remains essential when considering gravity as a gauge theory.\\

    Novel classes of modified gravity theories have emerged, incorporating curvature, torsion, and nonmetricity scalars. These classes arise even though the unmodified theories are mathematically equivalent at the equation level. The key lies in the difference between the torsion scalar $T$ and the nonmetricity scalar $Q$, which deviates from the usual Levi-Civita Ricci scalar $\m R$ of GR due to additional terms: $\m R = - T + B$ and $\m R = Q + B$, respectively; where $B$ is the boundary term. All quantities denoted by $(\, \m {}\,)$ are computed in relation to the Levi-Civita connection $\m \Gamma$. The objects framed in GR can be identified by an over-circle symbol. A geometric trinity of gravity of second-order can be observed in $\m R$, $B-T$, $Q+B$, whereas $f(\m R)$, $f(B-T)$, $f(Q+B)$ can be regarded as a geometric trinity of gravity of fourth-order \citep{Jimenez_2019_5, Capozziello_2023_83, Capozziello_2022_82}. Consequently, arbitrary functions $f(\m R)$, $f(T)$, and $f(Q)$ no longer share a total derivative relationship. Furthermore, scalar fields can be introduced within this framework, leading to theories of scalar-tensor \citep{Felice_2011_84, Horndeski_1974_10}, scalar torsion \citep{Bahamonde_2019_100_064018, Bahamonde_2023_107, Geng_2011_704}, and scalar nonmetricity \citep{Runkla_2018_98, Jarv_2018_97}, each offering intriguing possibilities. Recently, \citet{Heisenberg_2023.2309.15958} reviewed various cosmological models in $f(Q)$ gravity. Considering energy conditions, \citet{Banerjee_2021_81} investigated wormhole geometry in $f(Q)$ gravity. Several $f(Q)$ parameterizations have been analyzed, including observational constraints \citep{Narawade_2023_535} and investigating compact objects beyond the standard maximum mass limit \citep{Maurya_2023_269, Maurya_2022_2022_003, Lohakare_2023_526, Maurya_2022_70_2200061}. In addition, \citet{Boehmer_2023_2303.04463}, \citet{Palianthanasis_2024_43, Paliathanasis_2021_2402.02919} and \citet{Khyllep_2023_107} presented a dynamical system analysis in $f(Q)$ gravity with perturbations.\\

    In cosmology, the cosmic chronometer (CC) method is utilized to determine the age and expansion rate of the Universe. The CC technique consists of three primary components: (i) defining a sample of optimal CC tracers, (ii) determining the differential age, and (iii) assessing systematic effects \citep{Moresco_2022_25}. The Hubble parameter $H(z)$ is essential in determining the energy content of the Universe and its acceleration mechanism. The $H(z)$ estimation is mainly carried out at $z = 0$. Still, there are other methods to determine $H(z)$, such as the detection of baryonic acoustic oscillation (BAO) signal in the clustering of galaxies and quasars and analyzing Type Ia Supernovae (SNe Ia) observation \citep{Riess_2021_908, Font_Ribera_2014_2014_027, Raichoor_2020_500, Hou_2020_500, Riess_2018_853}. Pantheon$^+$ is an analysis that expands the original Pantheon framework to combine an even larger number of SNe Ia samples to understand the complete expansion history. In this study, we used the observational Hubble data (CC sample), Pantheon$^+$, and BAO data sets to investigate the expansion history of the Universe and the behavior of other geometrical parameters. \\
    
    This study investigates a specific subclass of the $f(Q, B)$ model to assess its potential as an alternative to the conventional cosmological framework. We have developed a numerical approach to predict the redshift behavior of the Hubble expansion rate. Our findings indicate that while the model can replicate the low-redshift behavior of the standard $\Lambda$CDM model, it exhibits notable differences at high redshifts. The $f(Q,B)$ model emerges as a viable candidate for explaining the current epochs and effectively captures the evolution of energy components over cosmic time, thereby supporting its validity as an alternative explanation for the observed acceleration of the Universe. We examined the background cosmological dynamics of the selected model and evaluated its feasibility using Bayesian analysis, supported by Markov Chain Monte Carlo (MCMC) methods, applied to late-time cosmic observations, including the Pantheon$^+$, CC and BAO data sets. Additionally, we introduced a dynamical system analysis to assess the stability of the model. A significant outcome of our analysis is the identification of a stable critical point within the dynamical system using center manifold theory (CMT). This critical point corresponds to the de Sitter phase, a well-established cosmological epoch characterized by accelerated expansion. The stability of this critical point suggests that, given certain initial conditions, the Universe will inevitably move toward and remain within the de Sitter phase. This finding aligns with current observations suggesting a late-time Universe dominated by DE and undergoing accelerated expansion.\\
    
    In teleparallel gravity, the boundary term $B$ can be incorporated into the Lagrangian, resulting in $f(T, B)$ theories that exhibit rich phenomenology \citep{Bahamonde_2015_92, Kadam_2023_83}. However, within the framework of nonmetricity gravity, the Lagrangian of symmetric teleparallel gravity does not account for the role of $B$. This has led to the development of the $f(Q, B)$ theory, which is currently of significant interest to cosmologists \citep{Capozziello_2023_83, De_2024_2024_50}. Our study explores the concept of an accelerating Universe by introducing a novel and straightforward parametrization for the Hubble parameter. This article is divided into five sections. Section \ref{Sec 2} presents the geometrical framework of symmetric teleparallelism, also formulating $f(Q,B)$ gravity and extracting the general metric and affine connection field equations. In Section \ref{Sec 3}, we apply this formulation to a cosmological setup, resulting in $f(Q, B)$ cosmology with observational data sets. Building on the model presented in Section \ref{Sec 4}, we perform a dynamical system analysis to investigate its long-term behavior and identify any stable or unstable states. Finally, Section \ref{Sec 5} concludes the article with the results and discussions.

\section{Symmetric teleparallel gravity} \label{Sec 2}
        We examine a gravitational model defined by the four-dimensional metric tensor $g_{\mu\nu}$ and the covariant derivative $\m \nabla_{\mu}~$, which is constructed using the generic connection $\Gamma_{\mu\nu}^{\zeta}\,$. Within the framework of Symmetric Teleparallel General Relativity (STGR), the connection $\Gamma_{\mu\nu}^{\zeta}$ is both flat and torsionless. Consequently, this results in $R_{;\eta\mu\nu}^{\zeta} = 0$ and $\mathrm{T}_{\mu\nu}^{\eta}=0$. Furthermore, it retains the symmetries of the metric tensor $g_{\mu\nu}$. Autoparallels are defined as \citep{Obukhov_2021_104}
    \begin{equation}
        \frac{d^{2}x^{\mu}}{ds^{2}}+\Gamma_{\zeta\nu}^{\mu}\frac{dx^{\zeta}}{ds}\frac{dx^{\nu}}{ds}=0 \, . \label{eq: 1}
    \end{equation}
        The nature of the geometry is determined by the affine connection, which is represented by $\Gamma_{\mu\nu}^{\zeta}$. 

    The three independent components can be expressed in the general form of an affine connection as follows:
        \begin{equation}\label{eq: affine connection}
            \Gamma^\sigma_{\mu \nu}=\m \Gamma_{\mu\nu}^{\sigma}+K^\sigma_{\,\,\,\mu\nu}+ L^\sigma_{\,\,\mu\nu}\, ,
        \end{equation}
    where $\m \Gamma_{\mu\nu}^{\sigma}$, $L^\sigma_{\,\,\mu\nu}$, and $K^\sigma_{\,\,\,\mu\nu}$ respectively represent the Levi-Civita connection, the deformation tensor, and the contortion tensor. These are defined as
        \begin{eqnarray} \label{connections formulae}
            \m \Gamma_{\mu\nu}^{\sigma}&=&\frac{1}{2}g^{ \sigma\zeta}\left(\partial_\mu g_{\zeta \nu}+\partial_ \nu g_{\zeta \mu}-\partial_\zeta g_{\mu \nu}\right)\, , \\
            L^\sigma_{\,\,\,\mu \nu}&=&\frac{1}{2}g^{\sigma \zeta} (-Q_{\mu \zeta \nu} - Q_{\nu \zeta \mu} + Q_{\zeta \mu \nu}) \, , \\ K^\sigma_{\,\,\,\mu\nu}&=&\frac{1}{2}g^{\sigma \zeta} (T_{\mu \zeta \nu} + T_{\nu \zeta \mu} + T_{\zeta \mu \nu})\, .
        \end{eqnarray}

    The Riemann tensor can be defined for the general connection
    \begin{equation}
        R_{\;\eta\mu\nu}^{\zeta} = \frac{\partial\Gamma_{\;\eta\nu}^{\zeta}% 
        }{\partial x^{\mu}}-\frac{\partial\Gamma_{\;\eta\mu}^{\zeta}}{\partial x^{\nu}}+\Gamma_{\;\eta\nu}^{\sigma}\Gamma_{\;\mu\sigma}^{\zeta}%
        -\Gamma_{\;\eta\mu}^{\sigma}\Gamma_{\;\nu\sigma}^{\zeta}\, , \label{eq: 2}
    \end{equation}
        the torsion tensor
    \begin{equation}
        \mathrm{T}_{\mu\nu}^{\eta}=\Gamma_{\;\mu\nu}^{\eta}-\Gamma_{\;\nu\mu}^{\eta}\, , \label{eq: 3}
    \end{equation}
        and the nonmetricity tensor 
    \begin{eqnarray}
        Q_{\eta\mu\nu}=\m \nabla_{\eta}g_{\mu\nu}=\frac{\partial g_{\mu\nu}%
        }{\partial x^{\eta}}-\Gamma_{\;\eta\mu}^{\sigma}g_{\sigma\nu}%
        -\Gamma_{\;\eta\nu}^{\sigma}g_{\mu\sigma}\, . \label{eq: 4}
    \end{eqnarray}        
        
        In symmetric teleparallel theory, we are always allowed to choose a suitable diffeomorphism that vanishes the general affine connection $\Gamma_{\mu\nu}^{\zeta}$, known as the coincident gauge \citep{Jimenez_2018_98}. As a consequence, the covariant derivative reduces to the partial derivative, and the symmetric teleparallel postulates (i.e., vanishing curvature and torsion) enforce that the general connection becomes the Levi-Civita connection, which is symmetric by construction.
    
        However, in the Teleparallel Equivalent of General Relativity (TEGR), the antisymmetric Weitzenb\"ock connection replaces the connection $\Gamma_{\mu\nu}^{\zeta}$. This results in $R_{;\eta\mu\nu}^{\zeta} = 0$ and $Q_{\eta\mu\nu}=0$. In this context, the torsion scalar $T$ becomes the fundamental geometric object in teleparallel gravity.

        As a result, the nonmetricity scalar $Q$, which is defined in \citep{Nester_1999}, has been introduced:
    \begin{equation}
        Q=Q_{\eta\mu\nu}P^{\eta\mu\nu} \, , \label{eq: 5}
    \end{equation}
        This statement represents the fundamental geometric quantity of gravity. The nonmetricity conjugate $P_{\;\mu\nu}^{\eta}$ is defined as
    \begin{equation}
        P_{\;\mu\nu}^{\eta}=-\frac{1}{4}Q_{\;\mu\nu}^{\eta}+\frac{1}{2}        Q_{(\mu\phantom{\eta}\nu)}^{\phantom{(\mu}\eta\phantom{\nu)}}+\frac
        {1}{4}\left(  Q^{\eta}-\tilde{Q}^{\eta}\right)  g_{\mu\nu}-\frac{1}
        {4}\delta_{\;(\mu}^{\eta}Q_{\nu)} \, , \label{eq: 6}
\end{equation}
        and the traces $Q_{\mu}=Q_{\mu\nu}^{\phantom{\mu\nu}\nu}$ and $\tilde{Q}_{\mu}=Q_{\phantom{\nu}\mu\nu}^{\nu\phantom{\mu}\phantom{\mu}}$ are used in this context. 
        
        The boundary term is defined as
\begin{equation}
        B = \mathring{R} - Q = -\mathring{\nabla}_\mu(Q^\mu-\tilde Q^\mu) = -\frac1{\sqrt{-g}}\partial_\mu\left[\sqrt{-g}(Q^\mu-\tilde Q^\mu)\right]\, .
\end{equation}
        The Ricci scalar $R$ corresponds to the Levi-Civita connection $\m{\Gamma}_{\mu\nu}^{\zeta}$ of the metric tensor $g_{\mu\nu}$. The nonmetricity scalar $Q$ for a symmetric and flat connection differs from $\m R$ by a boundary term $B$, which is defined as $B = \m R - Q$.

        The gravitational action integral for STGR is expressed as follows:
    \begin{equation}
        \int d^{4}x\sqrt{-g}Q\simeq\int d^{4}x\sqrt{-g}R - B \, , \label{eq: 7}
    \end{equation}
        this implies that STGR is dynamically equivalent to GR.

        However, the equivalence is lost when nonlinear components of the nonmetricity scalar $Q$ are introduced as in $f(Q)$--gravity in the gravitational action. Moreover, the corresponding gravitational theory no longer has dynamical equivalence with GR or its generalization, $f(\m R)$ gravity.

        The action integral for symmetric teleparallel $f(Q)$ gravity \citep{Jimenez_2018_98, Jimenez_2020_101} is expressed as follows:
    \begin{equation}
        S_{f\left(  Q\right)  }=\int d^{4}x\sqrt{-g}f(Q) \, . \label{eq: 8}
    \end{equation}

\subsection{\texorpdfstring{$f(Q,B)$}{} cosmology} \label{Sec 2a}

        A recent extension of the $f(Q,B)$ theory \citep{Capozziello_2023_83, De_2024_2024_50, Palianthanasis_2024_43} incorporates a boundary term into the gravitational action integral. This generalization includes the gravitational action integral in the following manner:
    \begin{equation} \label{eq: 9}
        S = \int d^4x \sqrt{-g} \left[ \frac{1}{2 \kappa} f(Q, B) + \mathcal{L}_m \right],
    \end{equation}
        where $g$ represents the determinant of the metric tensor $g_{\mu \nu}$, $\kappa = 8\pi G_N=1$, and $G_N$ denotes the gravitational constant. 
        
        To construct a realistic cosmological model, we consider a matter action $S_m$, associated with the energy-momentum tensor $\Theta_{\mu\nu}$. As shown in \citep{De_2024_2024_50}, varying the total action $S$ leads to the following Friedmann equations:
    \begin{eqnarray}
        \kappa T_{\mu\nu}=-\frac f2g_{\mu\nu}
        +\frac2{\sqrt{-g}}\partial_\eta \left(\sqrt{-g}f_Q P^\eta{}_{\mu\nu}\right)
        +(P_{\mu\alpha\beta}Q_\nu{}^{\alpha\beta}-2P_{\alpha\beta\nu}Q^{\alpha\beta}{}_\mu)
        f_Q
        +\left(\frac B2 g_{\mu\nu}-\m\nabla_{\mu}\m\nabla_{\nu}
    +g_{\mu\nu}\m\nabla^\alpha\m\nabla_\alpha-2P^\eta{}_{\mu\nu}\partial_\eta \right)f_B\,,\nonumber\\
        \label{eqn:FE1-pre}
    \end{eqnarray}
        This can be expressed in a covariant manner:
    \begin{eqnarray}
        \kappa T_{\mu\nu}=-\frac f2g_{\mu\nu}+2P^\eta{}_{\mu\nu} \m \nabla_\eta(f_Q-f_B)
        + \left(\m G_{\mu\nu}+\frac Q2g_{\mu\nu}\right)f_Q
        +\left(\frac B2g_{\mu\nu}-\m\nabla_{\mu}\m\nabla_{\nu} +g_{\mu\nu}\m\nabla^\alpha\m\nabla_\alpha \right)f_B\,. \label{eq: 11}
    \end{eqnarray}
        A definition of the effective stress-energy tensor is as follows
    \begin{eqnarray} \label{T^eff}
        T^{\text{eff}}_{\mu\nu} =  T_{\mu\nu}+ \frac 1{\kappa}\left[\frac f2g_{\mu\nu}-2P^\eta{}_{\mu\nu} \m \nabla_\eta(f_Q-f_B)-\frac {Qf_Q}2g_{\mu\nu} -\left(\frac B2g_{\mu\nu}-\m\nabla_{\mu}\m\nabla_{\nu} +g_{\mu\nu}\m\nabla^\alpha\m\nabla_\alpha \right)f_B\right]\,.
    \end{eqnarray}
        In order to produce an equation that is similar to that of GR
    \begin{eqnarray}
        \m G_{\mu\nu}=\frac{\kappa}{f_Q}T^{\text{eff}}_{\mu\nu}\,. \label{eq: 13}
    \end{eqnarray}
    In this section, we explore the application of $f(Q, B)$ gravity within a cosmological context and introduce $f(Q, B)$ cosmology. Our analysis focuses on a homogeneous and isotropic flat Friedmann--Lema\'itre--Robertson--Walker (FLRW) spacetime described by the line element in Cartesian coordinates
    \begin{eqnarray}
        ds^{2}=-dt^{2}+a^{2}(t)[dx^{2} + dy^{2}+dz^{2}],  \label{eq: 14}
    \end{eqnarray}
    where the scale factor is represented by $a(t)$, and it is related to the Hubble parameter through its first-time derivative, which is given by $H(t) = \frac{\dot{a}(t)}{a(t)}$.

    Following this section, it has been demonstrated that within the context of $f(Q, B)$ gravity, an additional effective sector of geometrical origin can be obtained as shown in Equation (\ref{T^eff}). Consequently, when considered in a cosmological context, this term can be interpreted as an effective DE sector, which possesses an energy-momentum tensor
    \begin{eqnarray}
            T^{\text{DE}}_{\mu\nu}= \frac 1{f_Q}\left[\frac 
            f2g_{\mu\nu}-2P^\eta{}_{\mu\nu}\m \nabla_\eta(f_Q-f_B)
            -\frac {Qf_Q}2g_{\mu\nu} -\left(\frac B2g_{\mu\nu}-\mathring{\nabla}_{\mu}\mathring{\nabla}_{\nu} + g_{\mu\nu} \mathring{\nabla}^\alpha \mathring{\nabla}_\alpha \right)f_B\right]\, , \label{eq: 15}
    \end{eqnarray}

    \begin{eqnarray}
        \mathring{R} =  6(2H^2 + \dot{H}), \quad Q = -6H^2,
        \quad B = 6(3H^2 + \dot{H}). \label{eq: 16}
    \end{eqnarray}
    In this case, we consider a vanishing affine connection $(\Gamma_{\mu \nu}^{\eta}=0)$, when fixing the coincident gauge. Our Friedmann-like equations can be derived from these data as follows:
    \begin{eqnarray}
            3 H^2 &=&\kappa \left(\rho_{\text{m}}+\rho_{\text{r}} + \rho_{\text{DE}}\right)\, ,  \label{first_field_equation}\\
            -2 \dot{H}-3 H^2& =&\kappa \left(\frac{\rho_{\text{r}}}{3} + p_{\text{DE}}\right) \, , \label{second_field_equation}
    \end{eqnarray}
    where $\rho_{\text{m}}$, $\rho_{\text{r}}$, $\rho_{\text{DE}}$ and $p_{\text{DE}}$ denote the matter density, radiation density, DE density, and DE pressure, respectively, treated as a perfect fluid. In the absence of interactions between nonrelativistic matter and radiation, each component independently adheres to its respective conservation laws, expressed as $\dot{\rho}_{\text{m}} + 3 H \rho_{\text{m}} = 0$ for matter and $\dot{\rho}_{\text{r}} + 4 H \rho_{\text{r}} = 0$ for radiation. Additionally, we have defined the effective DE density and pressure as follows:
        \begin{equation} \label{eq: 17}
            \rho_{\text{DE}}=\frac{1}{\kappa}\left[3 H^2\left(1-2 f_Q\right)-\frac{f}{2}+\left(9 H^2+3 \dot{H}\right) f_B-3 H \dot{f_B}\right],
        \end{equation}
        \begin{eqnarray} \label{eq: 18}
            p_{\text{DE}}=\frac{1}{\kappa}\Big[-2 \dot{H}\left(1-f_Q\right)-3 H^2\left(1-2 f_Q\right)+\frac{f}{2}+2 H \dot{f}_Q -\left(9 H^2+3 \dot{H}\right) f_B+\ddot{f_B}\Big].
        \end{eqnarray}

    Since standard matter is conserved independently, it can be deduced from Equations \eqref{eq: 17} and \eqref{eq: 18} that the DE density and pressure conform to the standard evolution equation
    \begin{eqnarray}
        \dot{\rho}_{\text{DE}}+3 H\left(\rho_{\text{DE}}+p_{\text{DE}}\right)=0 .
    \end{eqnarray}
    Finally, we can define the parameter for the DE equation of state (EoS) as
    \begin{eqnarray}
        \omega_{\text{DE}} = \frac{p_{\text{DE}}}{\rho_{\text{DE}}}.
    \end{eqnarray}

\subsection{Power law \texorpdfstring{$f(Q, B)$}{}} \label{SEC: 2b}
    In this study, we propose a specific mathematical form of $f(Q, B)$ to capture the characteristic power-law behaviors observed in different stages of the evolution history of the Universe, i.e., at different cosmological epochs. This form is inspired by the work of \citet{Bahamonde_2017_77_2}, which utilizes the Noether symmetry approach. The proposed form is given by:
    \begin{equation}
        f(Q, B) = f_0 Q^m B^n \, , \label{eq: fqb}
    \end{equation}    
    where $f_0$, $m$ and $n$ are arbitrary constants.

     To determine the theoretical values of the Hubble rate, we can numerically solve Equation \eqref{first_field_equation}. Assuming matter behaves as a pressureless perfect fluid ($p_{\text{m}}=0$), the matter density can be expressed as $\rho_\text{m} = 3 H_0^2 \Omega_{\text{m}0} (1+z)^3$, where $z$ denotes the cosmological redshift; it is defined as $\frac{a}{a_0} = \frac{1}{1 + z}$, where $a_0$ is the scale factor at present and $\Omega_{\text{m}0}$ is the current matter density parameter. Consequently, for the specific model under consideration, the first Friedmann equation can be written as follows:
    \begin{equation}
        \begin{aligned}
        H^{\prime \prime}(z) &=& \frac{-1}{f_0\, n (n-1) (1+z)^2 H(z)^3} \left[-9 f_0 (n-2 m-1) H(z)^4+f_0\, n (n-1) (1+z)\right. H(z)^3 H^{\prime}(z)\\
        & &-6 f_0 \left((n-1)^2+(2+n) m\right) (1+z) H(z)^3 H^{\prime}(z)+f_0\, n (n-1) (1+z)^2 H(z)^2 H^{\prime}(z)^2 \\
        & &+f_0 (1-n+2 m (1+n))(1+z)^2  H(z)^2 H^{\prime}(z)^2-6^{1-n-m} H_0^2(1+z)^3 \Omega_{\text{m} 0}\left(H(z)^2\right)^{-m} \\
        & & \left.\left(H(z)\left(3 H(z)-(1+z) H^{\prime}(z)\right)\right)^{2-n}\right] , \label{HZ_ode_QB}
        \end{aligned}
    \end{equation}
    where the prime $(')$ denotes differentiation with respect to $z$.

    In this study, we examine a second-order nonlinear ordinary differential equation that presents significant challenges for analytical solutions. As a result, we depend to numerical methods to solve Equation \eqref{HZ_ode_QB} in order to obtain theoretical values for the Hubble rate. To solve this equation, we need to apply suitable boundary conditions. The first boundary condition is straightforward: $H(0) = H_0$, which sets the present value of the Hubble parameter. To satisfy the second boundary condition, it is essential to confirm that the current rate of change of the Hubble parameter aligns with the projections of the standard $\Lambda$CDM model. This model describes the expansion of the Universe and provides a specific expansion law that the derivative should follow. By aligning the first derivative of $H(z)$ with this expansion law, we can accurately determine the second initial condition needed to solve the differential equation:
    \begin{eqnarray}
            H_{\Lambda \text{CDM}} = H_0 \sqrt{1 - \Omega_{\text{m}0} + \Omega_{\text{m}0} (1+z)^3} , \label{eq: lcdm}
    \end{eqnarray}
    After taking the derivative of the Equation \eqref{eq: lcdm} with respect to $z$, we can derive the second initial condition for Equation \eqref{HZ_ode_QB} as $H'(0) = \frac{3}{2}H_0 \Omega_{\text{m}0}$.

\section{Observational data, methodology and constraints} \label{Sec 3}

    To model the Universe accurately, we require robust observational data and effective parameter estimation methodologies. Within this framework, we detail the observational data sets and methods used to constrain the model parameters $f_0$, $m$, and $n$. Our analysis includes a comprehensive array of data, such as CC, Pantheon$^+$ and BAO from SNe Ia observations. We conducted MCMC analyses by integrating results from the CC, Pantheon$^+$, and BAO data sets, assuming a correlation among them. Our findings are presented for the CC + Pantheon$^+$ and CC + Pantheon$^+$ + BAO combinations, thereby reflecting their interdependence. By leveraging these diverse data sets, we effectively narrow down the model parameters, facilitating an in-depth exploration of the evolution of the Universe. Additionally, we explore $f(Q,B)$ gravity and its solutions involving the Hubble parameter. The CC data set, known for its reliability and model independence, provides Hubble parameters by measuring the age difference between two passively evolving galaxies. This method allows us to derive the Hubble function at various redshifts up to $z \approx 2$. The shape of $H(z)$ is further constrained by multiple sources, including 32 data points from Hubble data sets, BAO data from various sources, and CMB data from Planck 2018. The employed methodology, utilized data, and outcomes are detailed in subsequent sections.

\subsection{Cosmic Chronometers (CCs)} \label{Sec 3a}

    The Hubble parameter $H(z)$ can be estimated at certain redshifts $z$ using the following formula:
    \begin{equation}
        H(z) = \frac{\dot{a}}{a} = -\frac{1}{1+z}\frac{dz}{dt} \approx -\frac{1}{1+z}\frac{\Delta z}{\Delta t} \,  \label{eq: 20}
    \end{equation}

        where $\dot{a}$ is the derivative of the scale factor $a$ with respect to time $t$, and $\Delta z$ and $\Delta t$ are the differences in redshift and time, respectively, between two objects. The value of $\Delta z$ can be determined via a spectroscopic survey, while the differential ages $\Delta t$ of passively evolving galaxies can be used to estimate the value of $H(z)$. Compiling such observations can be regarded as a CC sample. We use 32 objects covering the redshift range $0.07 \leq z \leq 1.965$ \citep{Moresco_2022_25, Lohakare_2023_40_CQG}. For these measurements, one can construct a $\chi^2_{\text{CC}}$ estimator as follows:
    \begin{eqnarray}
        \chi^2_{CC} = \sum_{i=1}^{32} \frac{[H_{\text{th}}(z_i) - H_{\text{obs}}(z_i)]^2}{\sigma^2_{H}(z_i)} \, , \label{eq: 21}
    \end{eqnarray}
        where $H_{\text{obs}}$ and $H_{\text{th}}$ represent the observational and theoretical Hubble parameter values at redshift $z_i$, respectively, with $\sigma_H$ being the error in the observational value.

\subsection{Supernovae type Ia (SNe Ia)} \label{Sec 3b}

    We will also consider the Pantheon$^+$ SNe Ia data compilation, consisting of 1701 SNe Ia relative luminosity distance measurements spanning the redshift range of $0.00122 < z < 2.2613$ \citep{Brout_2022_938}. The Pantheon$^+$ data set contains distance moduli estimated from 1701 light curves of 1550 spectroscopically confirmed SNe Ia with a redshift range acquired from 18 distinct surveys. Notably, 77 of the 1701 light curves are associated with Cepheid-containing galaxies. The Pantheon$^+$ data set has the benefit of being able to constrain $H_0$ in addition to the model parameters. To fit the parameter of the model from the Pantheon$^+$ samples, we minimize the $\chi^2$ function. To calculate the chi-square $(\chi^2_{\text{SNe}})$ value, which helps estimate the best-fit parameters using SNe from the Pantheon compilation consisting of 1701 data points, we use the following formula:
    \begin{equation}
        \chi^2_{\text{SNe}}= \Delta\mu^T (C_{\text{Sys}+\text{Stat}}^{-1})\Delta\mu \, , \label{eq: 22}
    \end{equation}
        where $C_{\text{Sys}+\text{Stat}}^{-1}$ is the covariance matrix of the Pantheon$^+$ data set, which includes systematic and statistical uncertainties. As defined below, $\Delta \mu$ denotes the distance residual
    \begin{equation}
        \Delta\mu = \mu_{\text{th}}(z_i,\theta)-\mu_{\text{obs}}(z_i) \, . \label{eq: 23}
    \end{equation}
        \begin{figure} [htbp]
        \centering
        \includegraphics[width=110mm]{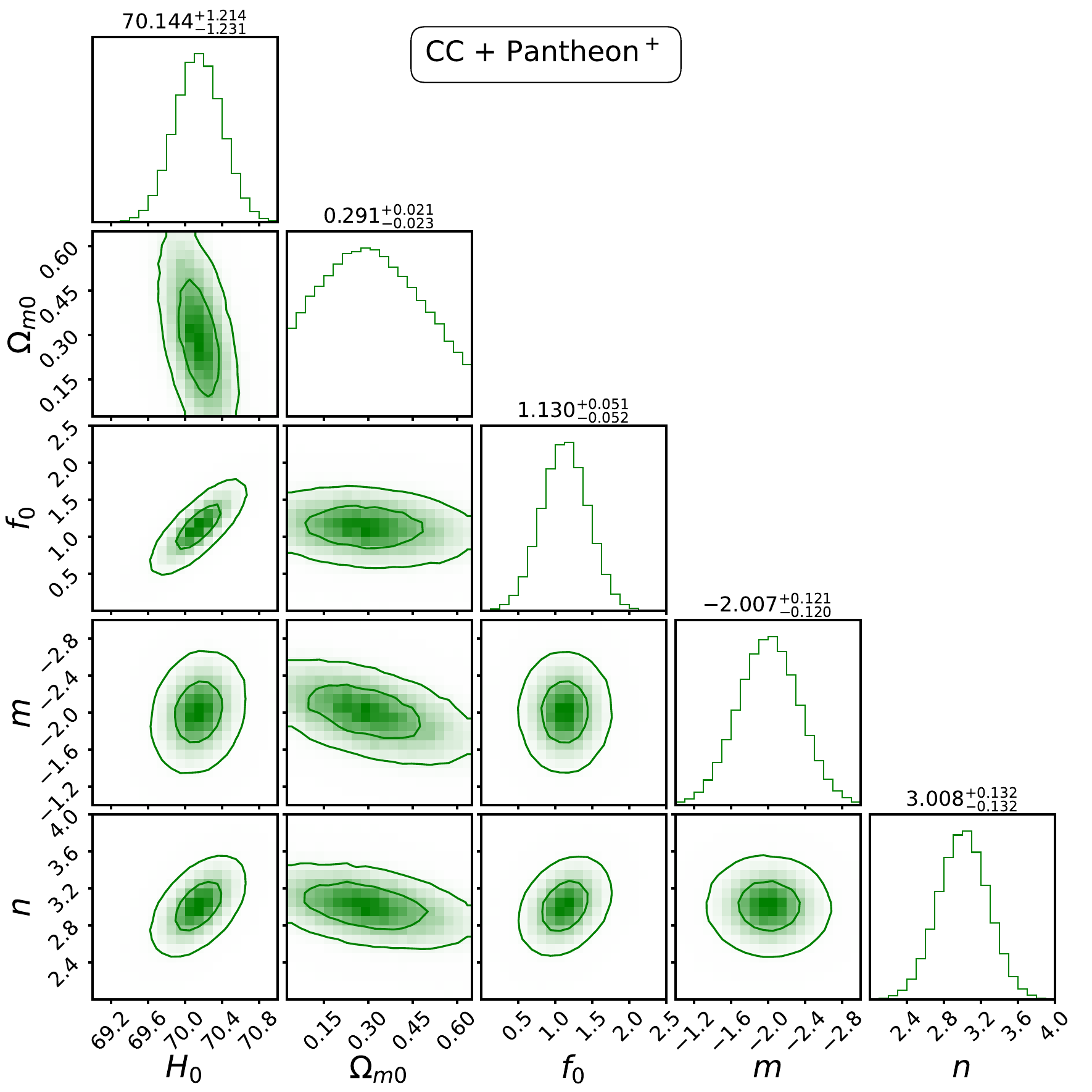}
        \caption{The contour plots display the $1\sigma$ and $2\sigma$ uncertainty regions for the model parameters $H_0$, $\Omega_{\text{m}0}$, $f_0$, $m$ and $n$. These contours are based on the combined CC + Pantheon$^+$ data sets.}
        \label{FIG1}
    \end{figure}

    \begin{figure} [htbp]
        \centering
        \includegraphics[width=110mm]{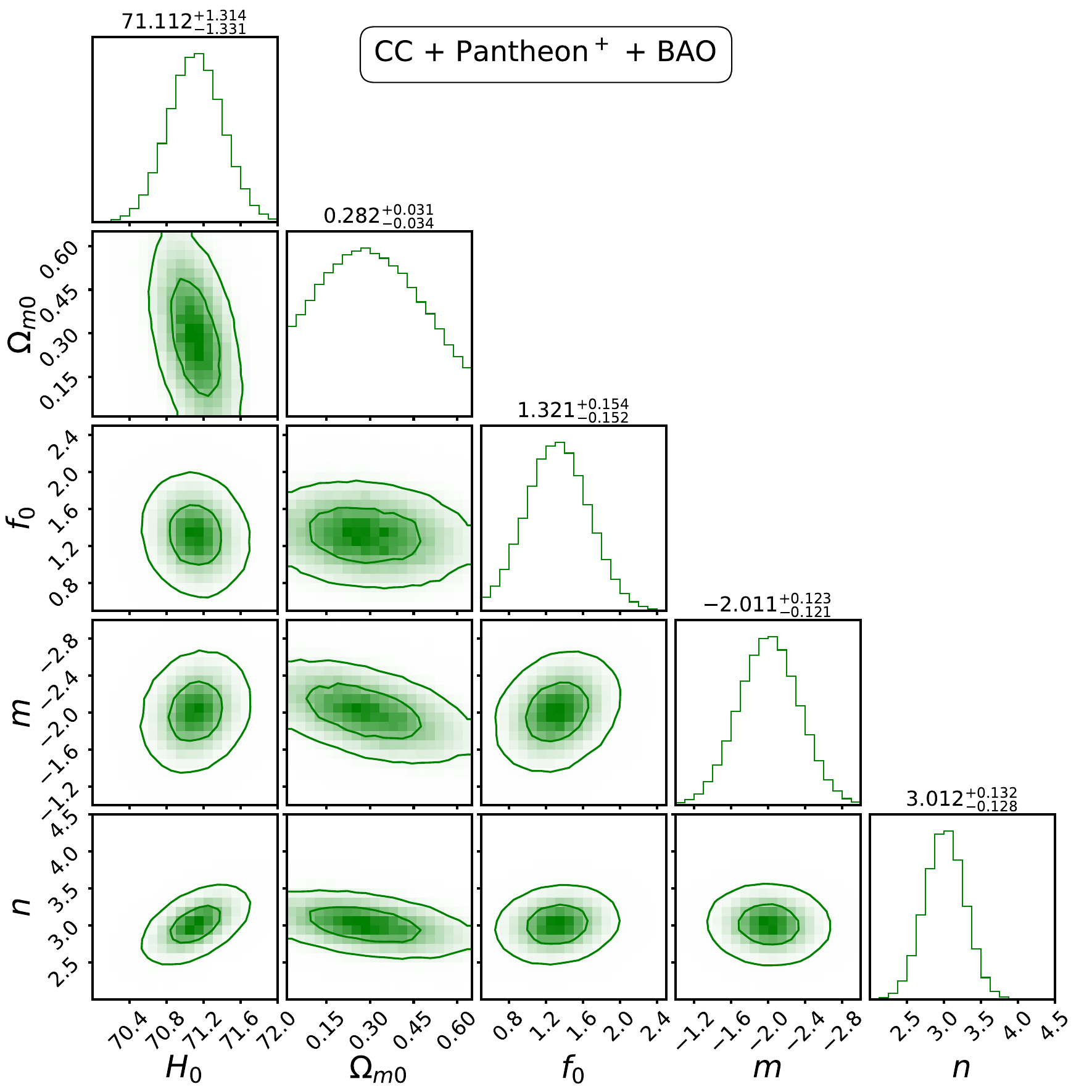}
        \caption{The contour plots display the $1\sigma$ and $2\sigma$ uncertainty regions for the model parameters $H_0$, $\Omega_{\text{m}0}$, $f_0$, $m$ and $n$. These contours are based on the combined CC + Pantheon$^+$ + BAO data sets.}
        \label{FIG2}
    \end{figure}
    Specifically, the distance modulus is defined as the difference between an observed apparent magnitude of an object, $m$, and its absolute magnitude, $M$, which measures its intrinsic brightness. At redshift $z_i$, the distance modulus is expressed as follows:
    \begin{equation}
        \mu_{\text{th}}(z_i,\theta)=5\log_{10}\left(d_L(z,\theta) \right) + 25 = m - M \, , \label{eq: 24}
    \end{equation}
        where $d_L$ represents the luminosity distance in Mpc depending on the model, which is
    \begin{equation}
        d_L(z_i,\theta)=\frac{c(1+z)}{H_0}\int_0^z \frac{d\zeta}{E(\zeta)}\, , \label{eq: 25}
    \end{equation}
        where $E(z)=\frac{H(z)}{H_0}$ and $c$ stands for the speed of light. Additionally, the residual distance is denoted by
    \begin{equation}
        \Delta\Bar{\mu}=\begin{cases}
            \mu_k-\mu_k^{cd}, & \text{if $k$ is in Cepheid hosts}\\
            \mu_k-\mu_{\text{th}}(z_k), & \text{otherwise}
        \end{cases} \label{eq: 26}
    \end{equation}
        where $\mu_k^{cd}$ is the Cepheid host-galaxy distance that SH0ES revealed. This covariance matrix can be coupled with the SNe covariance matrix to construct the covariance matrix for the Cepheid host galaxy. Equipped with statistical and systematic uncertainties from the Pantheon$^+$ data set, the combined covariance matrix is represented as $C^{\text{SNe}}_{\text{Sys}+\text{Stat}}+C^{cd}_{\text{Sys}+\text{Stat}}$. The expression above defines the $\chi^2$ function for the combined covariance matrix used to constrain cosmological models in the analysis:
    \begin{equation}
        \chi^2_{\text{SNe}^+}= \Delta\Bar{\mu} (C^{\text{SNe}}_{\text{Sys}+\text{Stat}}+C^{cd}_{\text{Sys}+\text{Stat}})^{-1}\Delta\Bar{\mu}^T \, . \label{eq: 27}
    \end{equation}

        \begin{figure} [htbp]
        \centering
        \includegraphics[width=.48\textwidth]{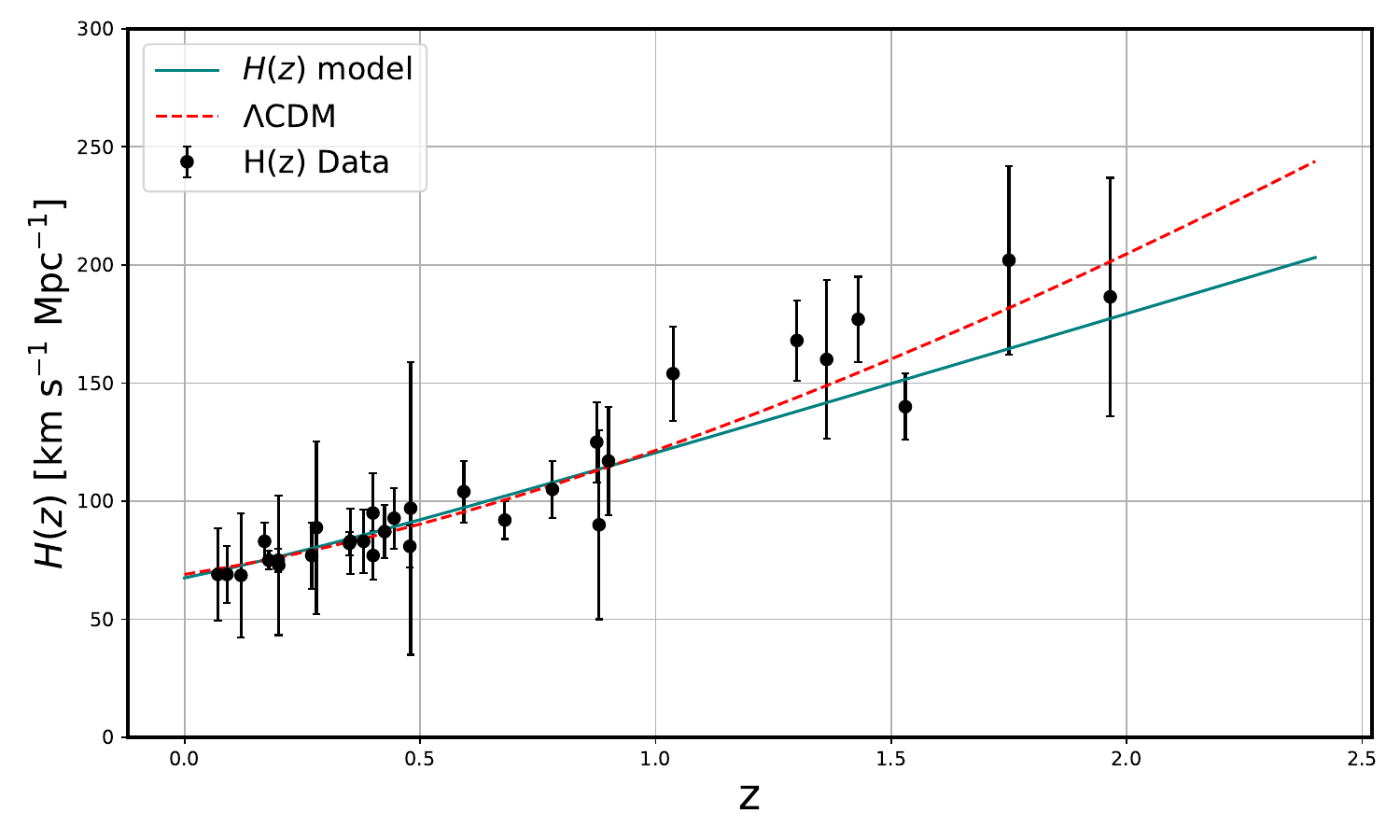}
        \includegraphics[width=.48\textwidth]{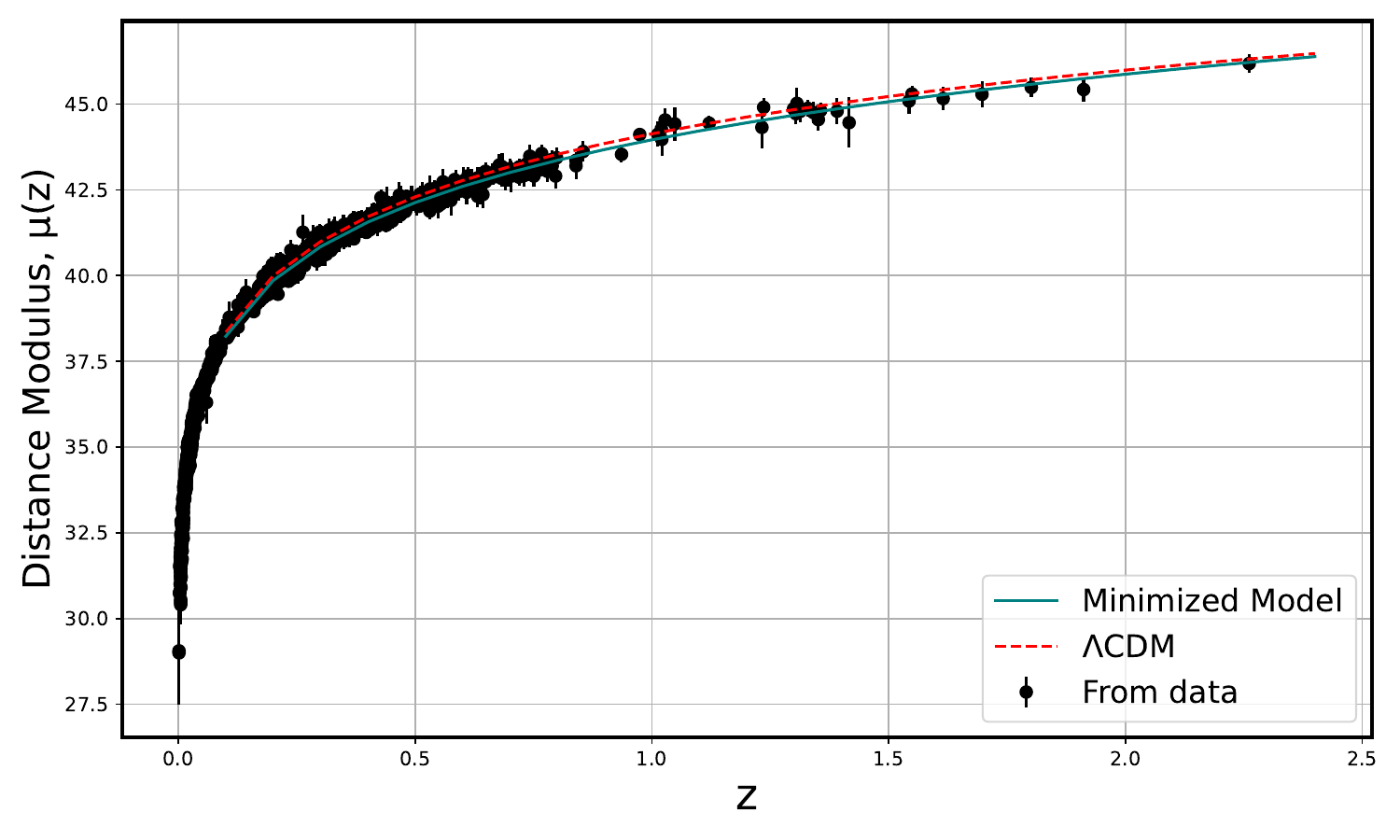}
        \caption{In the left panel, the black error bars represent the uncertainty associated with the 32 data points from the CC sample. The solid teal line corresponds to the model, while the dashed red line represents the $\Lambda$CDM. Moving to the right panel, we observe a red line that depicts the plot of the distance modulus of model $\mu(z)$ against redshift $z$. The teal line demonstrates a superior fit to the 1701 data points from the Pantheon$^+$ data set, including their associated error bars.}
        \label{FIG3}, 
    \end{figure}

    \begin{figure*} [htbp]
        \centering
        \includegraphics[width=180mm]{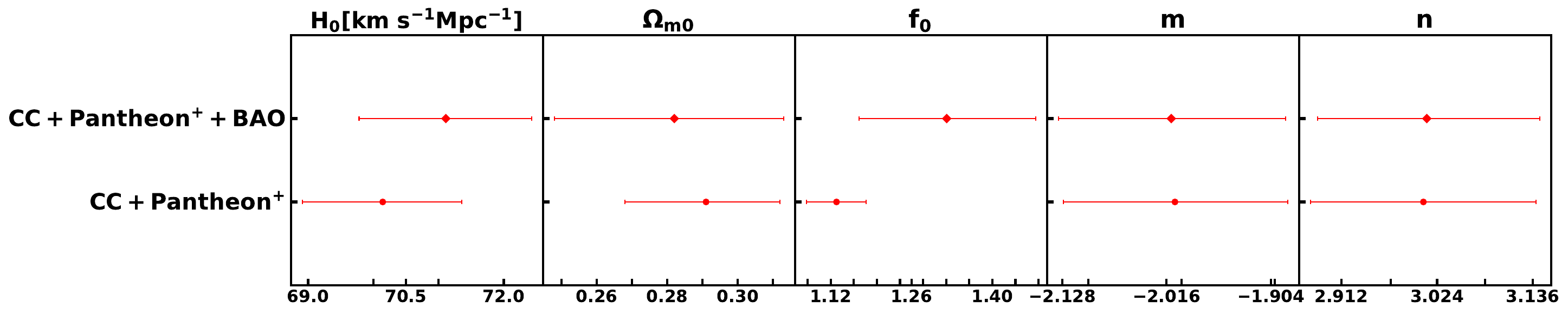}
        \caption{A whisker plot showing the model parameters $H_0$, $\Omega_{\text{m}0}$, $f_0$, $m$, and $n$, highlighting their discrepancies.}
        \label{FIG: whisker_plot}, 
    \end{figure*}

\subsection{Baryon Acoustic Oscillation (BAO)} \label{Sec 3c}
    Various surveys have significantly advanced the analysis of cosmic distance scales through BAO, including the 6-degree Field Galaxy Survey, Sloan Digital Sky Survey, and WiggleZ Dark Energy Survey. These surveys have measured BAO signals at multiple redshifts, providing critical data for understanding the expansion of the Universe. The horizon of sound during the radiation drag era, denoted as $r_s(z_*)$, is calculated using the following equation:
    \begin{equation}
        r_s\left(z_*\right)=\frac{c}{\sqrt{3}} \int_0^{\frac{1}{1+z_*}} \frac{\left(a^2 H\right)^{-1} d a}{\sqrt{1+\left(3 \Omega_{b 0} / 4 \Omega_{\gamma 0}\right) a}} \, ,
    \end{equation}
    where $c$ represents the speed of light and $\Omega_{b0}$ and $\Omega_{\gamma 0}$ are the present baryon and photon densities, respectively.
    
    To derive BAO constraints, the angular diameter distance $D_A(z)$, dilation scale $D_v(z)$ and the Hubble parameter $H(z)$ are used, defined by the equations
    \begin{equation}
        \begin{gathered}
        d_A(z)=\int_0^z \frac{d z^{\prime}}{H\left(z^{\prime}\right)} \, , \\
        D_v(z)=\left(\frac{d_A(z)^2 c z}{H(z)}\right)^{1 / 3} \, .
        \end{gathered}
    \end{equation}
    
    For the MCMC analysis, the same walkers, steps, and priors as in the CC sample are utilized. The chi-square function for BAO is expressed as
        \begin{equation}
            \chi_{\text{BAO}}^2 = X^T C^{-1} X \, ,
        \end{equation}
    where $X$ and $C^{-1}$ are the data vector and the inverse covariance matrix, respectively defined in Ref. \citep{Giostri_2012_2012_027}.    
    
    \begin{figure} [htbp]
    \vspace{0.5cm}
        \centering
        \includegraphics[width=.48\textwidth]{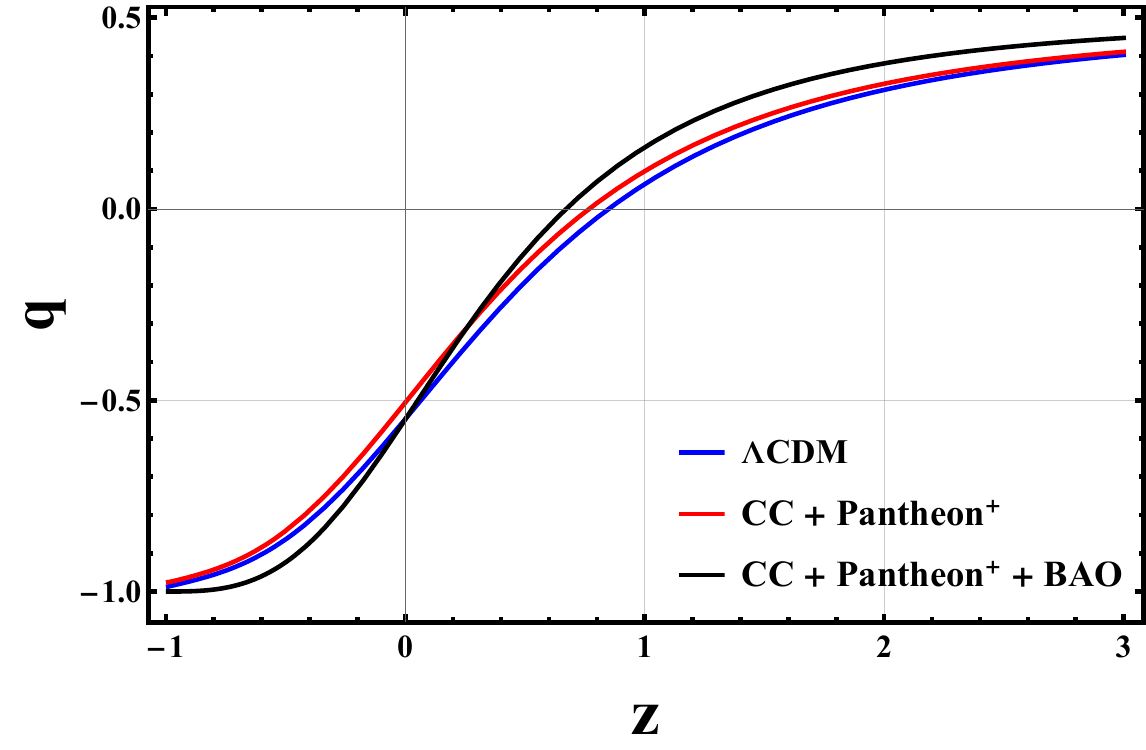}
        \includegraphics[width=.48\textwidth]{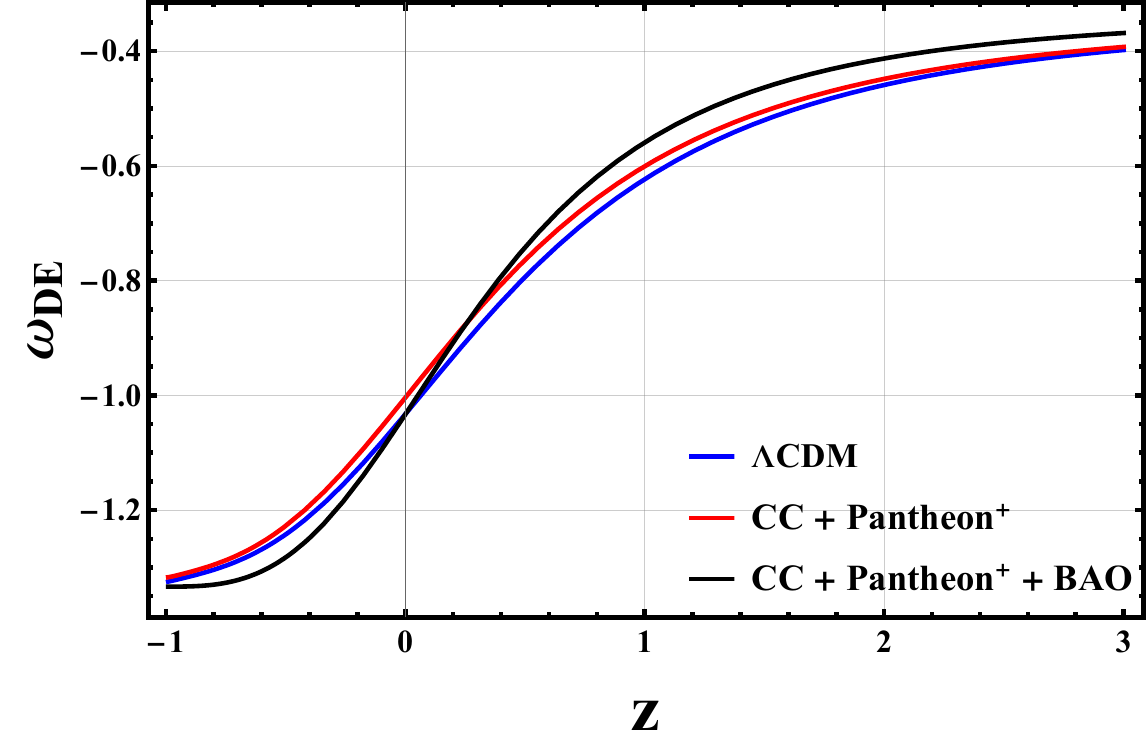}
        \caption{Behavior of the deceleration parameter (left panel) and EoS parameter (right panel) using the CC + Pantheon$^+$ and CC + Pantheon$^+$ + BAO data sets, with the mean values of parameters $ f_0 $, $ m $, and $ n $ as listed in Table \ref{TABLE I}.}
        \label{FIG4}
    \end{figure}
       
    \begin{table*} [htbp]
        \centering
    \begin{tabular}{|*{6}{c|}}\hline
        {\centering  \textbf{Data sets}} & $H_0$ & $\Omega_{\text{m}0}$ & $f_0$ & $m$ & $n$ \\ [0.5ex]
    \hline \hline
        \parbox[c][0.7cm]{3cm}{\centering \textbf{CC + Pantheon$^+$}} & $70.144^{+1.214}_{-1.231}$ & $0.291^{+0.021}_{-0.023}$ & $1.130^{+0.051}_{-0.052}$ & $-2.007^{+0.121}_{-0.120}$ & $3.008^{+0.132}_{-0.132}$  \\
    \hline
        \parbox[c][0.7cm]{3.5cm}{\centering \textbf{CC + Pantheon$^+$ + BAO}} & $71.112^{+1.314}_{-1.331}$ & $0.282^{+0.031}_{-0.034}$ & $1.321^{+0.154}_{-0.152}$ & $-2.011^{+0.123}_{-0.121}$ & $3.012^{+0.132}_{-0.128}$ \\[0.5ex] % [1ex] 
    \hline
    \end{tabular}
        \caption{Above table that shows the parameters explored by the MCMC algorithm. The table includes the parameter name, the corresponding best-fit values. An overview of the $H_0$, $\Omega_{\text{m}0}$, $f_0$, $m$ and $n$ parameter findings from the MCMC study, which is based on the CC, Pantheon$^+$, and BAO data sets.}
        \label{TABLE I}
    \end{table*}

        \begin{table*}[htbp]
        \centering
        \scalebox{0.72}{
        \begin{tabular}{|*{12}{c|} }
    \hline
        \parbox[c][0.7cm]{3.2cm}{\centering  \textbf{Data sets}}  &\multicolumn{2}{c|}{$\chi^2_{\text{min}}$} &\multicolumn{2}{c|}{AIC} &\multicolumn{2}{c|}{AIC$_\text{c}$} &\multicolumn{2}{c|}{BIC} & {$\Delta \text{AIC}$} & {$\Delta \text{AIC}_\text{c}$} & {$\Delta \text{BIC}$} \\
    \cline{2-9}
        \parbox[c][0.7cm]{1.5cm}& $f(Q, B)$ & $\Lambda$CDM & $f(Q, B)$ & $\Lambda$CDM & $f(Q, B)$ & $\Lambda$CDM & $f(Q, B)$ & $\Lambda$CDM & & &\\ 
    \hline \hline
        \parbox[c][0.7cm]{3cm}{\centering \textbf{CC + Pantheon$^+$}} & 1652.231 & 1654.270 & 1662.231 & 1658.270 & 1662.265 & 1658.277 & 1668.421 & 1660.746 & 3.961 & 3.988 & 7.675 \\
    \hline
        \parbox[c][0.7cm]{3.5cm}{\centering \textbf{CC + Pantheon$^+$ + BAO}} & 1659.321 & 1659.123 & 1669.321 & 1663.123 & 1669.355 & 1663.129 & 1675.521 & 1665.603 & 6.198 & 6.226 & 9.918 \\
    \hline
    \end{tabular}}
        \caption{The table presents the minimum $\chi^2$ values for the $f(Q, B)$ model, along with their corresponding AIC, AIC$_\text{c}$, and BIC values. It also includes a comparison of the differences with $\Lambda$CDM model in AIC, AIC$_\text{c}$, and BIC values.}
        \label{TABLE I b}
    \end{table*}

    Figures \ref{FIG1} and \ref{FIG2} provide the contour plots with $1\sigma$ and $2\sigma$ errors for the CC + Pantheon$^+$ and CC + Pantheon$^+$ + BAO data sets, respectively. The values obtained from the MCMC analysis that best fit the model are shown in Table \ref{TABLE I}. In the left panel of Figure \ref{FIG3}, the evolution of the Hubble parameter as a function of redshift is illustrated. This figure compares the predictions of two models: the $\Lambda$CDM model and the $H(z)$ model derived from the numerical approach proposed in this study (depicted by the teal line), alongside observational data. One line is included for the $\Lambda$CDM model to facilitate the comparison. Additionally, the dashed red line (labeled as $\Lambda$CDM) is derived from the standard prediction of the $\Lambda$CDM model with the parameters $H_0$, and $\Omega_\text{m}$. In Figure \ref{FIG3}, the right panel shows a comparison of the distance modulus using our $f(Q,B)$ model (teal line) and the $\Lambda$CDM model (dashed red line) predictions. Both models were considered with their respective parameters. The similarity between our model and the $\Lambda$CDM prediction is evident. However, when the same shared parameter values were used, the models deviated from each other, mainly in the apparent magnitude prediction. The dashed red line in Figure \ref{FIG3} shows that our model fits better with the $\Lambda$CDM. Also, the model accurately captures the behavior of the Hubble function, as shown by the consistency of the error bars. In Figure \ref{FIG3}, the observed distance modulus of the 1701 SNe Ia data set is depicted, along with the best-fit theoretical curves of the distance modulus function $\mu (z)$ shown as a teal line.

    Figure \ref{FIG: whisker_plot} displays the best-fit values and associated uncertainties for the model parameters $H_0$, $\Omega_{\text{m}0}$, $f_0$, $m$, and $n$, derived from our MCMC analysis. The plot visually represents the parameter ranges obtained from different data sets, including CC + Pantheon$^+$ and CC + Pantheon$^+$ + BAO. The Hubble constant $H_0$ values range from approximately 68.913 $\text{Km} \, \, \text{s}^{-1} \, \text{Mpc}^{-1}$ to 72.426 $\text{Km} \, \, \text{s}^{-1} \, \text{Mpc}^{-1}$, while the matter density parameter $\Omega_{\text{m}0}$ spans from 0.248 to 0.312. The parameters $f_0$, $m$, and $n$ exhibit ranges of 1.078 to 1.475, $-2.127$ to $-1.89$, and 2.876 to 3.144, respectively. These ranges highlight the variability and discrepancies in the parameter estimates, underscoring the robustness and reliability of the model fits to the observational data. The whisker plot effectively conveys the uncertainties inherent in the model parameters, providing a comprehensive overview of the results from the MCMC analysis. This comparative analysis not only enhances our understanding of the data but also facilitates informed decision-making based on observed patterns.
 
    Figure \ref{FIG4} illustrates the significance of the deceleration parameter $q$, a crucial metric in cosmology that provides insights into the dynamics of the Universe. A positive $q$ indicates deceleration, while a negative $q$ signifies acceleration. Analysis of the CC + Pantheon$^+$ and CC + Pantheon$^+$ + BAO data sets reveals that $q$ transitions from positive in the past, indicating early deceleration, to negative in the present, indicating current acceleration, as depicted in Figure \ref{FIG4}. At the current cosmic epoch, the deceleration parameter $q_0$ has been measured as $-0.506$ and $-0.549$ for the CC + Pantheon$^+$ and CC + Pantheon$^+$ + BAO data sets, respectively. These values are in good agreement with the range of $q_0 = -0.528^{+0.092}_{-0.088}$ determined by recent observations \citep{Christine_2014_89}. Current observations align with this deceleration parameter, and the derived model demonstrates a smooth transition from deceleration to acceleration at $z_t = 0.763$ and $ z_t = 0.67 $ for the CC + Pantheon$^+$ and CC + Pantheon$^+$ + BAO data sets, respectively. The recovered transition redshift $z_t$ is consistent with current constraints based on 11 $H(z)$ observations reported by \citet{Busca_2013_552} for redshifts $0.2 \leq z \leq 2.3$, $z_t = 0.74 \pm 0.5$ from \citet{Farooq_2013_766}, $z_t = 0.7679^{+0.1831}_{-0.1829}$ by \citet{Capozziello_2014_90_044016}, and $z_t = 0.60^{+0.21}_{-0.12}$ by \citet{Yang_2020_2020_059}. Similarly, the EoS parameter $(\omega_{\text{DE}})$, is integral to understanding the evolution of the Universe, as it correlates with the energy sources influencing this progression. The current EoS values for DE, represented by $\omega_{\text{DE}}(z = 0)$, are determined to be $-1.032$ and $-1.004$ for the CC + Pantheon$^+$ and CC + Pantheon$^+$ + BAO data sets, respectively. Various cosmological studies have also placed constraints on the EoS parameter. For instance, the Planck 2018 results yielded $\omega_{\text{DE}}=-1.03\pm 0.03$ \citep{Aghanim_2020_641}, and the WAMP + CMB analysis reported $\omega_{\text{DE}}=-1.079^{+0.090}_{-0.089}$ \citep{Hinshaw_2013_208}. By computing the associated energy density and pressure of DE, we can observe the fluctuations in the effective DE EoS, which are depicted in redshift (Figure \ref{FIG4}).

    We evaluate the models against the standard $\Lambda$CDM model using the Akaike Information Criterion (AIC) and the Bayesian Information Criterion (BIC), in addition to $\chi^2_{\text{min}}$. Both the AIC and BIC consider the model's goodness of fit and its complexity, which depends on the number of parameters $(n)$. The AIC is calculated as
    \begin{eqnarray}
            \text{AIC} = \chi^2_{\text{min}} + 2 n \, ,
    \end{eqnarray}
In statistical modeling, a lower AIC value indicates a better fit to the data, accounting for model complexity. This penalizes models with more parameters, even if they fit the data better. The BIC is computed as
    \begin{eqnarray}
            \text{BIC} = \chi^2_{\text{min}} + n \, \text{ln} \,\mathcal{N} \, ,
    \end{eqnarray}
where $ \mathcal{N} $ is the number of data samples used in the MCMC process. The corrected AIC (AIC$_\text{c}$) is defined as
    \begin{eqnarray}
    \text{AIC}_\text{c} = \text{AIC} + \frac{2 n (n+1)}{\mathcal{N}-n-1} \, ,
    \end{eqnarray}
for large sample sizes $ (\mathcal{N} \gg n) $, the correction term becomes negligible, making AIC$_\text{c}$ preferable over the original AIC.

We compare the AIC and BIC values between the $ f(Q, B) $ model and the $\Lambda$CDM model to gain insights into how well each model aligns with the standard cosmological model. The differences in AIC, AIC$_\text{c}$ and BIC are expressed as $ \Delta \text{IC} = \text{IC}_{\text{Model}} - \text{IC}_{\Lambda\text{CDM}} $. Smaller $ \Delta $AIC and $ \Delta $BIC values indicate that a model, along with its selected data set, closely resembles the $\Lambda$CDM model, suggesting superior performance. To assess the effectiveness of our MCMC analysis, we computed the corresponding AIC, AIC$_\text{c}$, and BIC values, as shown in Table \ref{TABLE I b}. Our results strongly endorse the proposed $ f(Q,B) $ gravity models based on the analyzed data sets. Additionally, we observed that the $ f(Q, B) $ model exhibits higher precision when applied to the CC + Pantheon$^+$ data sets.

\section{Dynamical System Analysis} \label{Sec 4}  
    
    The methods of dynamical systems are valuable for analyzing the overall long-term dynamics of a particular cosmological model. This involves an equation, $x' = f(x)$, where $x$ is a column vector and $f(x)$ is the equivalent vector of autonomous equations. In this method, the prime symbol represents the derivative with respect to the number of $e$--folding, $N = \text{ln} \, a(t)$. The general form of the dynamical system for the modified FLRW equations defined by Equation \eqref{first_field_equation} can be generated through this approach. Let us define a new variable: 
    \begin{eqnarray}
        X=f_B, \hspace{0.3cm} Y=\frac{\dot{f}_B}{H}, \hspace{0.3cm} Z=\frac{\dot{H}}{H^2}, \hspace{0.3cm} V=\frac{\kappa \, \rho_\text{r}} {3H^2}, \hspace{0.3cm} W=-\frac{f}{6 H^2}, \hspace{0.3cm} \Omega_{\text{m}}= \frac{\kappa \, \rho_{\text{m}}}{3 H^2}, \hspace{0.3cm} \Omega_{\text{DE}}=\frac{\kappa \, \rho_{\text{DE}}}{3 H^2} \, . \label{eq: 30} 
    \end{eqnarray}
    Thus, from Equation \eqref{first_field_equation}, we have the algebraic identity
    \begin{eqnarray}
        \Omega_{\text{m}} + \Omega_{\text{r}} + \Omega_{\text{DE}}=1 \, , \label{eq: 31}
    \end{eqnarray}
    together with the density parameters
    \begin{eqnarray}
        & & \Omega_{\text{m}} = \frac{\kappa \rho_\text{m}}{3 H^2}, \hspace{0.6cm} \Omega_{\text{r}} = V = \frac{\kappa \rho_r}{3 H^2},\\ \hspace{0.3cm} & &\Omega_{\text{DE}}= 1-2 f_Q + W + 3 X + X Z - Y \, . \label{eq: density parameters}
    \end{eqnarray}

    So, taking the derivative of these variables with respect to $N$, we obtain the following dynamical system:
    \begin{subequations} \label{eq: DS}   
    \begin{eqnarray} 
        \frac{dX}{dN} &=& Y \, ,\\
        \frac{dY}{dN} &=& 2 - 3V + \frac{2 Z}{3} - \frac{2 Z f_Q}{3} + 3 X - 2 f_Q + W -\frac{2 f_Q'}{3} + X Z - Y Z \, ,\\
        \frac{dZ}{dN} &=& \lambda - 2 Z^2 \, ,\\
        \frac{dV}{dN} &=& -4 V - 2 Z V \, ,
    \end{eqnarray}
    \end{subequations}
    where $\lambda = \frac{\ddot{H}}{H^3}$, we will concentrate on the scenario where $f(Q, B) = f_0 Q^m B^n$. The model for this scenario can be expressed using the dynamical variables
    \begin{eqnarray}
        f_Q = m \, W
    \end{eqnarray}
    and we get the following dependency relation:
    \begin{eqnarray}
        W &=& -\frac{X}{n} (Z+3) \, ,\\
        \Omega_{\text{m}} &=& -V-2 f_Q + W + 3 X + X Z - Y \, ,\\
        \Omega_{\text{DE}} &=& 1-2 f_Q + W + 3 X + X Z - Y \, .
    \end{eqnarray}
    and
    \begin{eqnarray}
        \lambda = \frac{Y (Z + 3)-2 X Z \left(m (Z + 3) + 3 (n-1)\right)}{(n-1) X} \, .
    \end{eqnarray}
    It is possible to eliminate the equations for $W$, $\Omega_{\text{m}}$ and $\Omega_{\text{DE}}$ from our autonomous system using the relations mentioned and Equation \eqref{eq: 31}, resulting in a set of only four equations:
    \begin{subequations} \label{eq: Autonomous_DS}   
    \begin{eqnarray} 
        \frac{dX}{dN} &=& Y \, ,\\
        \frac{dY}{dN} &=& \frac{1}{3 n (n-1)}\Big[n \big(-2 X (Z+3) (m (Z-3)+3) + 2 m Y (Z + 3) + 9 V + 3 Y Z - 2 (Z + 3)\big) \nonumber\\ & & - (2 m-1) X (Z + 3) (2 m Z + 3) + n^2 (-9 V + 3 X (Z + 3)-3 Y Z + 2 Z + 6)\Big] \, ,\nonumber\\ \\
        \frac{dZ}{dN} &=& \lambda - 2 Z^2 \, ,\\
        \frac{dV}{dN} &=& -4 V - 2 Z V \, .
    \end{eqnarray}
    \end{subequations}

        It is important to note that a dynamical system has a critical point, and this point must be taken into account when analyzing the system,
    \begin{eqnarray}
        \mathcal{P}_{\star}(X, Y, Z, V) = \Big( \frac{2 n}{3 (n-1)},\,  0,\, 0, \, 0\Big) \, ,
    \end{eqnarray}
        for the existence condition $m = 1 - n$. We will now analyze the range of values for $n$, which will result in a stable critical point. While we will not explicitly mention the area of instability, whether it is saddle-like or repulsor-like, it is important to note that the critical point is a de Sitter acceleration phase; therefore, any kind of instability of the critical point is not supported by observations.

        The eigenvalue is given by
        \begin{eqnarray} \label{eigenvalues}
        \left\{0,-4,\frac{-n-\sqrt{-2 n^2-3 n}}{n},\frac{- n+\sqrt{-2 n^2-3 n}}{n}\right\} \, .
        \end{eqnarray}

        The phase portrait in Figure \ref{FIG5} shows the behavior of a dynamical system near a stable critical point, $\mathcal{P}_{\star}$. As time progresses, trajectories in the phase space tend to move toward $\mathcal{P}_{\star}$, indicating that it is an attractor for the system. This convergence from various initial conditions signifies that small perturbations decay over time, returning the system to the stable state at $\mathcal{P}_{\star}$. The stability of $\mathcal{P}_{\star}$ can be analyzed using the Jacobian matrix evaluated at $\mathcal{P}_{\star}$. The stability of $\mathcal{P}_{\star}$ is ensured when all eigenvalues of the Jacobian possess negative real parts, which aligns with the observed behavior in the phase portrait. The overall dynamics of the system are governed by the differential equations defining it, and the phase portrait provides a graphical representation to visualize these dynamics and understand the long-term behavior of the system. The zero eigenvalues suggest the presence of a center manifold, which requires further analysis to determine overall stability.

\subsection{Stability analysis for \texorpdfstring{$\mathcal{P}_{\star}$}{} by using CMT}
    
    CMT is a crucial aspect of dynamical systems theory, focusing on the behavior of systems near fixed points. As detailed by \citet{Perko_2013}, CMT provides a framework for understanding the stability of these points, especially when traditional linear stability theory falls short due to the presence of zero eigenvalues. This theory reduces the dimensionality of the system near critical points, allowing for a more manageable analysis of stability. When a system approaches a critical point, it tends to behave according to an invariant local center manifold, denoted as $ W^c $. This manifold is associated with eigenvalues that have zero real parts, capturing the essential dynamics of the system near equilibrium.

    Consider a function $ f $ belonging to $ C^r(E) $, where $ E $ is an open subset of $ \mathbb{R}^n $ that includes the origin, and $ r \geq 1 $. Assume $ f(0) = 0 $ and that the derivative $ Df(0) $ possesses $ c $ eigenvalues with zero real parts and $ s $ eigenvalues with negative real parts, where $ c + s = n $. Generally, the system can be reformulated as follows:
    \begin{equation}\label{CMT_system}
        \begin{aligned}
        & \dot{x}=A x+F(x, y)\, , \\
        & \dot{y}=B y+G(x, y) \, ,
        \end{aligned}
    \end{equation}
    where $A$ is a square matrix with $c$ eigenvalues having zero real parts and $ B $ is a square matrix with $ s $ eigenvalues having negative real parts and $(x, y) \in \mathbb{R}^c \times \mathbb{R}^s$. The functions $ F $ and $ G $ satisfy $ F(0) = G(0) = 0 $ and their derivatives at zero are also zero. Furthermore, there is a small positive value $ \epsilon > 0 $ and a function $ g(x) $ in $ C^r(N_\epsilon(0)) $, which defines the local center manifold and satisfies certain conditions,
    \begin{equation}\label{local center manifold}
        \operatorname{Dg}(x)[A x+F(x, g(x))]-B g(x)-G(x, g(x))=\mathcal{N}(g(x))=0 \, ,
    \end{equation}
    for $ |x| < \epsilon $. The center manifold can be derived using the system of differential equations,
    \begin{equation}
        \dot{x}=A x+F(x, g(x)) ,
    \end{equation}
    for all $ x \in \mathbb{R}^c $ with $ |x| < \epsilon $.
    
    At the critical point $\mathcal{P}_{\star}$, the Jacobian matrix for the autonomous system represented by Equation \eqref{eq: Autonomous_DS} is as follows:
        \begin{equation}
            J\left(\mathcal{P}_{\star}\right)=\left(
            \begin{array}{cccc}
            0 & 1 & 0 & 0 \\
            -3+\frac{3}{n} & -2 & -\frac{4}{3} & -3 \\
            0 & \frac{9}{2 n} & 0 & 0 \\
            0 & 0 & 0 & -4 \\
            \end{array}\right)
        \end{equation}
    
    The eigenvalues of the Jacobian matrix, as presented in Equation \eqref{eigenvalues}, are $\lambda_1 = 0$, $\lambda_2 = -4$, $\lambda_3 = \frac{-n-\sqrt{-2 n^2-3 n}}{n}$, and $\lambda_4 = \frac{-n+\sqrt{-2 n^2-3 n}}{n}$. The corresponding eigenvectors are $[\frac{4 n}{9-9 n},0,1,0]^T$, $[-\frac{3 n}{11 n+3},\frac{12 n}{11 n+3},-\frac{27}{22 n+6},1]^T$, $[\frac{2 n}{9},\frac{2}{9} \left(\sqrt{-n (2 n+3)}-n\right),1,0]^T$ and $[\frac{2 n}{9},\frac{1}{9} (-2) \left(n+\sqrt{-n (2 n+3)}\right),1,0]^T$.\\

    Using the CMT, we examine the stability of the critical point $\mathcal{P}_{\star}$. By applying the transformation $X = x - \frac{2 n}{3 (n-1)}$, $Y = y$, $Z = z$, and $V = v$, we shift this critical point to the origin. The resulting equations in the new coordinate system are as follows:
    \begin{equation}
    \left(\begin{array}{c}
    \dot{x} \\
    \dot{y} \\
    \dot{z} \\
    \dot{v} 
    \end{array}\right)=\left(\begin{array}{cccc}
     0 & 0 & 0 & 0 \\
     0 & -4 & 0 & 0 \\
    0 & 0 & \frac{-n-\sqrt{-2 n^2-3 n}}{n} & 0 \\
    0 & 0 & 0 & \frac{-n+\sqrt{-2 n^2-3 n}}{n}  
    \end{array}\right)\left(\begin{array}{l}
    x \\
    y \\
    z \\
    v 
    \end{array}\right)+\left(\begin{array}{c}
    \text { non } \\
    \text { linear } \\
    \text { term }
    \end{array}\right)
    \end{equation}
    Upon examining the diagonal matrix in relation to the standard form (Equation \eqref{CMT_system}), it is clear that the variables $y$, $z$ and $v$ remain stable, while $x$ acts as the central variable. At this critical point, matrices $A$ and $B$ take on the following form:\\
        $$A=\left[0\right] , \quad
        B=\left[\begin{array}{ccc}
        -4 & 0 & 0  \\
        0 & \frac{-n-\sqrt{-2 n^2-3 n}}{n} & 0  \\
        0 & 0 & \frac{-n+\sqrt{-2 n^2-3 n}}{n} 
        \end{array}\right]  
        $$
    In the context of CMT, the manifold is characterized by a continuous differential function. Assuming specific functions for the stable variables $y = g_1 (x)$, $z = g_2 (x)$, and $v = g_3 (x)$, we derived the zeroth approximation of the manifold functions using Equation \eqref{local center manifold}
    \begin{eqnarray}
        & & N(g_1 (x)) = \left(-3+\frac{3}{n}\right) x+\mathcal{O}^2,\nonumber\\& & N(g_2 (x)) = 0 + \mathcal{O}^2 , \hspace{0.2cm} N(g_3 (x)) = 0 + \mathcal{O}^2,
    \end{eqnarray}
    where the $\mathcal{O}^2$ term encompasses all terms that are proportional to the square or higher powers. The following expression gives the center manifold in this scenario:
    \begin{eqnarray}
        \dot{x} = \left(-3+\frac{3}{n}\right) x + \mathcal{O}^2.
    \end{eqnarray}

    According to the CMT, the critical point $\mathcal{P}_{\star}$ exhibits stable behavior for $(n < 0) \vee (n > 1)$.

    \begin{figure} [htbp]
        \centering
        \includegraphics[width=85mm]{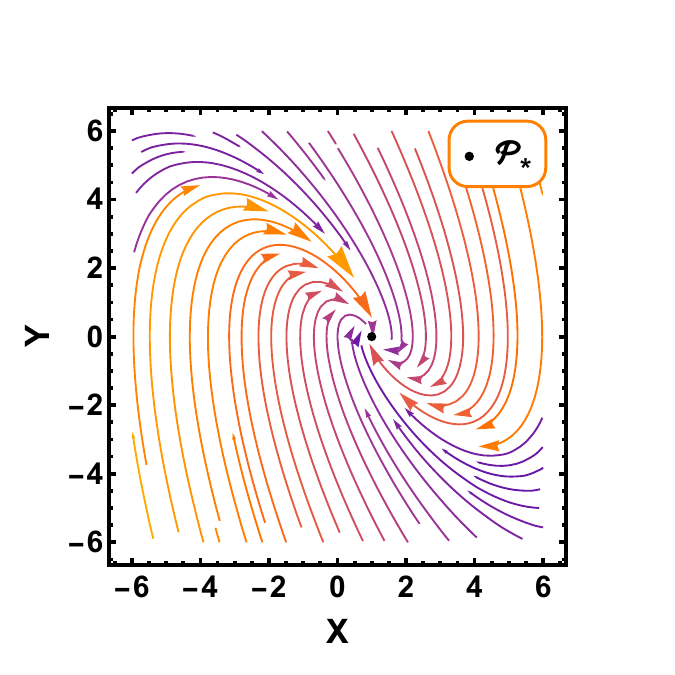}
        \caption{This graph shows the behavior of a four-dimensional system (Equation \eqref{eq: Autonomous_DS}) simplified to two dimensions. The parameters are set as $Z = 0$, $V = 0$, and $n = 3.012$.}
        \label{FIG5}
    \end{figure}

    Figure \ref{FIG5} reveals that a specific point in this two-dimensional representation (the critical point $\mathcal{P}_{\star}$) attracts other points over time, suggesting its stable and attractive nature within the full four-dimensional system. Figure \ref{FIG5} shows the fascinating world of sink trajectories within a dynamical system, visualized through a phase portrait. This point signifies a location where trajectories tend to sink or converge. The critical point $\mathcal{P}_{\star}$ is nonhyperbolic due to the presence of zero eigenvalues and can describe the acceleration of the Universe. It is also an attractor solution, stable in the regions $( n < 0 ) \vee ( n > 1 )$ as determined by CMT. The density parameters for radiation, matter, and DE are $\Omega_\text{r} = 0$, $\Omega_\text{m} = 2-\frac{4}{n-1}$, and $\Omega_\text{DE} = -1 + \frac{4}{n-1}$, respectively, satisfying the constraint in Equation \eqref{eq: 31}. This scenario corresponds to a deceleration parameter $q = -1$ and a total EoS $\omega_\text{tot} =-1-\frac{2 \dot{H}}{3 H^2}= -1$, indicating a de Sitter phase and, consequently, an accelerating expansion of the Universe.

    \begin{figure} [htbp]
        \centering
        \includegraphics[width=100mm]{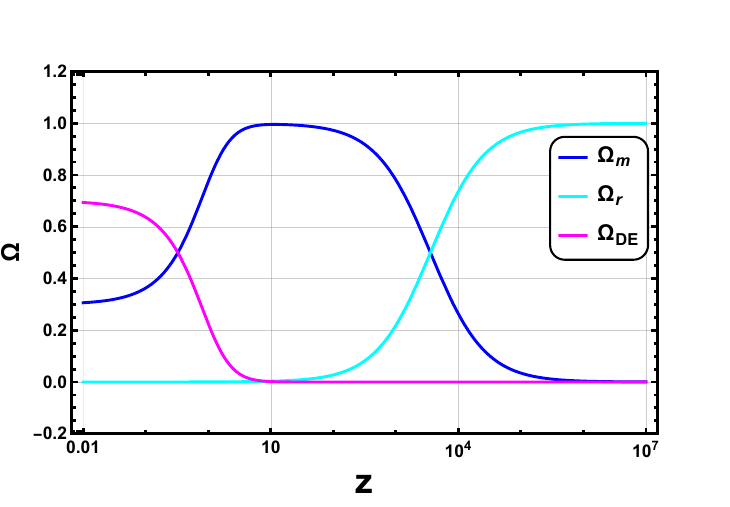}
        \caption{Evolution of density parameters DE (magenta line), matter (blue line), and radiation (cyan line) for the initial conditions: $X =10^{11}$, $Y = 3.29 \times 10^{14}$, $Z = 0.008$, and $V = 4.54 \times 10^{-5}$.}
        \label{FIG6}
    \end{figure}
    Figure \ref{FIG6} shows the evolution of density parameters for DE, matter, and radiation as a function of redshift. The model parameters $m = -2.011$ and $n = 3.012$, which are obtained from the parametrization method using MCMC analysis for the CC + Pantheon$^+$ + BAO data sets, are used.
    \begin{itemize}
        \item The \textit{magenta line} represents the DE density parameter ($\Omega_{\text{DE}}$), which increases sharply at lower $z$, indicating the growing influence of DE in the accelerated expansion of the Universe.
        \item The \textit{blue line} shows the matter density parameter ($\Omega_{\text{m}}$), which decreases with increasing $z$, reflecting the dilution of matter as the Universe expands.
        \item The radiation density parameter ($\Omega_{\text{r}}$) is represented by the \textit{cyan line}, which remains almost constant at zero for small redshift values. This emphasizes its minimal contribution in the present epoch.
    \end{itemize}

    In Figure \ref{FIG6}, we present the evolution of the density parameters for DE, matter, and radiation as a function of redshift ($z$), utilizing a model parameter value of $m=-2.011$ and $n = 3.012$ obtained from our MCMC analysis. The initial conditions for this plot are $X =10^{11}$, $Y = 3.29 \times 10^{14}$, $Z = 0.008$, and $V = 4.54 \times 10^{-5}$. The magenta line representing DE exhibits a significant increase at lower redshifts, indicating its dominance in the current epoch of the Universe. The blue line for matter density decreases as redshift decreases, reflecting the transition from a matter-dominated Universe at higher redshifts to a DE-dominated Universe at lower redshifts. The cyan line for radiation density is notably higher at early times (high redshifts) and diminishes rapidly as the Universe expands, consistent with the radiation-dominated era in the early Universe. This plot effectively captures the dynamic evolution of the energy components of the Universe, illustrating the transitions from radiation dominance to matter dominance and finally to DE dominance, providing valuable insights into the evolution of the Universe.

\section{Summary and Conclusion} \label{Sec 5}
    In this paper, we have delved into the cosmological implications of a modified $f(Q,B) $ gravity model, which integrates both the nonmetricity scalar $ Q $ and the boundary term $ B $. Our approach adopted the coincident gauge, where the general affine connection vanishes, meaning the covariant derivative reduces to the partial derivative. We then applied Bayesian statistical analysis using MCMC techniques to constrain the model parameters. The analysis was grounded in observational data from CC measurement, the extended Pantheon$^+$ data set, and BAO measurements. Our results elucidate a smooth transition from a deceleration phase to an accelerating expansion phase in the evolution of the Universe. This transition is critical in understanding the dynamics of cosmic expansion and the role of DE. We developed a numerical approach to predict the redshift behavior of the Hubble expansion rate. This approach was instrumental in constraining the model parameters and understanding the kinematic evolution of the Universe. The $ f(Q,B)$ model has been compared with the standard $ \Lambda $CDM model, demonstrating its potential as a viable alternative cosmological framework. While the $ \Lambda $CDM model has been the cornerstone of modern cosmology, our findings suggest that the $ f(Q, B) $ model can replicate the low-redshift behavior of the $ \Lambda $CDM model and exhibits notable differences at high redshifts. Our findings align strongly with current cosmological observations of a late-time Universe dominated by DE and undergoing accelerated expansion. This supports the validity of the $ f(Q, B) $ model as an alternative explanation for the observed acceleration of the Universe. Our study reveals a significant divergence in the $H_0$ measurements derived from the CC + Pantheon$^+$ and CC + Pantheon$^+$ + BAO data sets, as illustrated in Figure \ref{FIG: whisker_plot}. This discrepancy underscores the persistent tension surrounding the Hubble constant within the field of cosmology. Resolving these discrepancies is critical for improving our comprehension of the expansion rate of the Universe.\\

    A dynamical system analysis framework has been introduced to assess the stability of the model. The identification of a stable critical point using CMT underscores the robustness of the $ f(Q, B) $ model. A significant finding of our study is the identification of a stable critical point within the dynamical system of the model, corresponding to the de Sitter phase. The stability of this critical point implies that, given specific initial conditions, the Universe will inherently move toward and stay within the de Sitter phase. This observation aligns with current data indicating a Universe dominated by DE and undergoing late-time accelerated expansion. Future research could delve deeper into the specific initial conditions leading to the de Sitter phase and investigate the influence of the boundary term $ B $ on the dynamics of the system. The density parameter plot depicts the transition of the Universe from radiation dominance to matter dominance and ultimately to DE dominance. This offers valuable insights into the evolutionary dynamics of the Universe. Additionally, exploring the implications of this stable critical point for physical quantities like the Hubble parameter would offer valuable insights into the evolution of the Universe. In summary, the $ f(Q, B) $ gravity model not only aligns well with current cosmological observations but also provides a comprehensive framework for understanding the accelerated expansion of the Universe. The ability of the model to capture the transition from deceleration to acceleration, identify a stable critical point, and offer a viable alternative to the $ \Lambda $CDM model makes it a promising candidate for further exploration in cosmological studies. Our study emphasizes the crucial role of modified gravity theories in understanding the expansion of the Universe and the nature of DE.

\section*{Acknowledgements} S.V.L. would like to express gratitude for the financial support provided by the University Grants Commission (UGC) through the Senior Research Fellowship (UGC Reference No. 191620116597) to carry out the research work. B.M. would like to acknowledge the support given by IUCAA, Pune (India), through the visiting associateship program.

%\section*{References}



\end{document}